\documentclass[prl,twocolumn,aps,showpacs,10pt]{revtex4-1}

\usepackage{graphicx}  
\graphicspath{{./Figures/}}
\usepackage{dcolumn}  
\usepackage{amssymb, amsmath}
\usepackage{hyperref}
\usepackage{multirow}
\usepackage{color}

\begin{document}

\title{$^{3}$He-$^{129}$Xe Comagnetometery using $^{87}$Rb  Detection and Decoupling}
\author{M.\ E.\ Limes}
\author{D.\ Sheng}
\author{M.\ V.\ Romalis}
\affiliation{Department of Physics, Princeton University, Princeton, New Jersey, 08544, USA}
\date{\today}
\begin{abstract}
We describe a $^{3}$He-$^{129}$Xe comagnetometer using $^{87}$Rb atoms for noble-gas spin polarization and detection. We use a train of $^{87}$Rb $\pi$ pulses and $\sigma^+/\sigma^-$ optical pumping to realize a finite-field Rb magnetometer with suppression of spin-exchange relaxation. We suppress frequency shifts from polarized Rb by measuring the $^{3}$He and $^{129}$Xe spin precession frequencies in the dark, while applying  $\pi$ pulses along two directions to depolarize Rb atoms. The plane of the $\pi$ pulses is rotated to suppress the Bloch-Siegert shifts for the nuclear spins. We measure the ratio of $^{3}$He to $^{129}$Xe spin precession frequencies with sufficient absolute accuracy to resolve the Earth's rotation without changing the orientation of the comagnetometer. A frequency resolution of  7~nHz is achieved after integration for 8 hours without evidence of significant drift. 

\end{abstract}

\maketitle
%
Spin comagnetometers first introduced in \cite{Lamoreaux_1986} are used for several types of fundamental physics experiments, such as tests of Lorentz, CP and CPT symmetries \cite{Smiciklas,Rosenberry,Baker,Brown} and searches for spin-dependent forces \cite{Venema_1992,Vasilakis,Bulatowicz_2013,Tullney_2013, Hunter}. They also have practical applications as inertial rotation sensors \cite{Harle_1993,Donley_2010,Kitching_2011,Donley2013,Meyer_2014,Walker_2016}.
When two different spin ensembles occupy the same volume they experience nearly the same average magnetic field \cite{Sheng_2014}. The ratio of their spin precession frequencies  $f_r = \omega_{\text{He}}/\omega_{\text{Xe}}$ can then be used to measure the inertial rotation rate $\Omega$ or a  spin coupling beyond the Standard Model $b$:
\begin{equation}
  f_r =(\gamma_{\text{He}}B_0 + \Omega_{z}+ b_{z}^{\text{He}})/(\gamma_{\text{Xe}}B_0 +\Omega_z+ b_z^{\text{Xe}}).
  \label{eq:fr}
  \end{equation}
where $B_0$ is  the bias field along $\hat{z}$ and $\gamma_{\text{He}}$, $\gamma_{\text{Xe}}$ are the gyromagnetic ratios for $^{3}$He and $^{129}$Xe, which are well known  \cite{Makulski_2015}. Since $I=1/2$ nuclear spins are free from quadrupolar energy shifts \cite{Happer_1987}, $f_r$ provides an absolute measure of non-magnetic spin interactions---this is particularly important in searches for spin-gravity coupling  \cite{Flambaum_2009} (where the interaction is hard to modulate), and for use as a gyroscope.

An alkali-metal magnetometer provides a natural way to detect nuclear-spin signals because Rb atoms are already used to polarize the nuclear spins by spin-exchange collisions; these collisions enhance the classical dipolar field from the nuclear magnetization by a factor $\kappa_0$ \cite{Schaefer}, which is about 5 for Rb-$^{3}$He \cite{Romalis_1998} and 500 for Rb-Xe \cite{Ma_2011}. However, the presence of polarized Rb atoms also causes large noble-gas frequency shifts that affect the accuracy of Eq.~(\ref{eq:fr}). In the past, these frequency shifts have been avoided in $^{3}$He-$^{129}$Xe comagnetometers by detecting a smaller dipolar field outside of an alkali-free cell using an RF coil \cite{Chupp_1994} or a SQUID magnetometer \cite{Gemmel_2010}.

\begin{figure}[ht!]
\center
\includegraphics{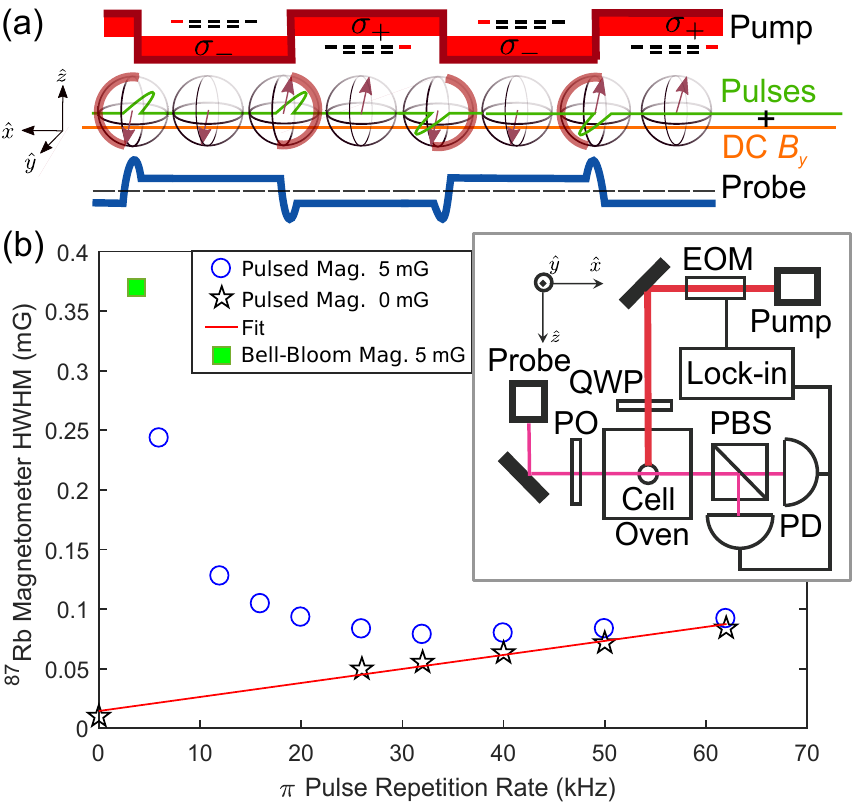}
\caption{ (a) The pulse-train $^{87}$Rb magnetometer uses $\hat{y}$-axis $\pi$ pulses with a $\sigma^{_+}$-$\sigma^{_-}$ pump beam along $\hat{z}$ and a linearly polarized probe beam along $\hat{x}$ (see inset). A $B_y$ field gives a paramagnetic Faraday rotation signal detected with a lock-in.
(b) Using a $^{87}$Rb-N$_2$ cell at $103^{\circ}$C and low laser power, we measure the magnetometer linewidth in  zero bias field (stars) and for $B_0 =5$~mG (circles) as a function of $\pi$ pulse repetition rate. For comparison we also show a Bell-Bloom magnetometer linewidth (square) that is limited by Rb-Rb spin-exchange relaxation.}
\label{fig:1}
\end{figure}

In this Letter we describe a new method for operating the $^{3}$He-$^{129}$Xe comagnetometer using \(^{87}\)Rb readout with high sensitivity and accuracy. We develop a \(^{87}\)Rb magnetometer that can operate in a finite magnetic field of about 5 mG while suppressing Rb-Rb spin-exchange relaxation to increase the magnetometer sensitivity.  It  uses a train of \(^{87}\)Rb $\pi$ pulses in unison with $\sigma^{_+}$/$\sigma^{_-}$-modulated optical pumping to refocus spin precession of \(^{87}\)Rb in the \(B_{0}\) field, while retaining sensitivity to transverse  fields (see Fig. \ref{fig:1}a). To suppress nuclear-spin frequency shifts we use a Ramsey pump-probe sequence where the $^{3}$He and $^{129}$Xe frequencies are measured in the absence of Rb laser light \cite{Sheng_2014}. In addition, we suppress back-polarization of Rb due to noble gases by a factor of $10^3$ using a train of $^{87}$Rb $\pi$ pulses along both $\hat{x}$ and $\hat{y}$ directions. Finally, we show that the effect of the  $^{87}$Rb $\pi$ pulses on the ratio of the noble gas frequencies can be eliminated by rotating the plane of the pulses about the $\hat{z}$ axis at the sum of the  $^{3}$He and $^{129}$Xe spin precession frequencies. Using this method and Eq.~(\ref{eq:fr}) we can measure the Earth's rotation without changing the orientation of the magnetometer, in contrast to previous measurements that required physical rotation of the apparatus to modulate the signal \cite{Brown}.
  
{\em Pulse-train Alkali Magnetometer.--}  The effect of $^{87}$Rb $\pi$ pulses can be understood by considering precession about $B_0$ of the $F_a=I+1/2$ and $F_b=I-1/2$ $^{87}$Rb spin manifolds, which have opposite gyromagnetic ratios $\pm\gamma_{\text{Rb}}$  \cite{Savukov_2005}. After a $\pi$ pulse, the  relative orientation of  $\left\langle \mathbf{F}_a \right\rangle$ and $\left\langle \mathbf{F}_b \right\rangle$ remains the same.
If the time  $\tau$ between $\pi$ pulses along $\hat{y}$ is small, $\tau\ll(\gamma_{\text{Rb}} B_0)^{-1}$, the spin evolution due to $B_0$ is cancelled to first order.  Thus, the magnetometer effectively operates at zero magnetic field in the spin-exchange relaxation free (SERF) regime \cite{Allred_2002}. The only limitation comes from the finite duration of the $\pi$ pulses, $t_p$, during which spin-exchange relaxation occurs. As shown in \cite{Korver_2013}, the remaining relaxation rate is proportional to the duty cycle of the pulses $t_p/\tau$.

To demonstrate spin-exchange suppression in the pulse-train $^{87}$Rb magnetometer we use a $^{87}$Rb-N$_2$ cell. The cell is heated in a boron nitride oven by AC currents and is located inside five cylindrical $\mu$-metal shields.  We use two sets of field coils, a larger set to provide the bias field  and a smaller set of three-axis square Helmholtz coils for $^{87}$Rb tipping pulses (3.5 in.~side, two turns, $3~\mu$H). The  $^{87}$Rb $\pi$ pulses (following sequence $\pi_y$$\pi_y$$\pi_{\text{-}y}$$\pi_{\text{-}y}$) are typically 1.5-3 $\mu$s in duration and 1-2 A in amplitude. They are generated with a pulsed current source using an LT1210 driver with an LC filter at the output to eliminate any DC currents in the coil.

In Fig.~\ref{fig:1}b we show measurements of the magnetic resonance linewidth for a $B_y$ field magnetometer signal. When the bias field $B_0=0$, the geometry matches a SERF magnetometer. The resonance linewidth increases linearly with the duty cycle of the $\pi$ pulses. When $B_0=5$~mG, the linewidth drops as the $\pi$ pulse rate increases, a result of canceling the precession of $\left\langle \mathbf{F}_a \right\rangle$ and $\left\langle \mathbf{F}_b \right\rangle$ away from each other due to the $B_0$ field. Simultaneously, the amplitude of response to a $B_y$ field increases \cite{supp}. Finally, we show the linewidth for a Bell-Bloom magnetometer, where  $\sigma^{_+}$/$\sigma^{_-}$ polarization is modulated at the $^{87}$Rb Larmor frequency (3.7 kHz) and the magnetometer signal corresponds to a change in $B_z$ field. This provides a measure of the spin-exchange-limited linewidth in our bias field, and demonstrates that our pulse train reduces the linewidth by a factor of 5.

\begin{figure}
\center
\includegraphics{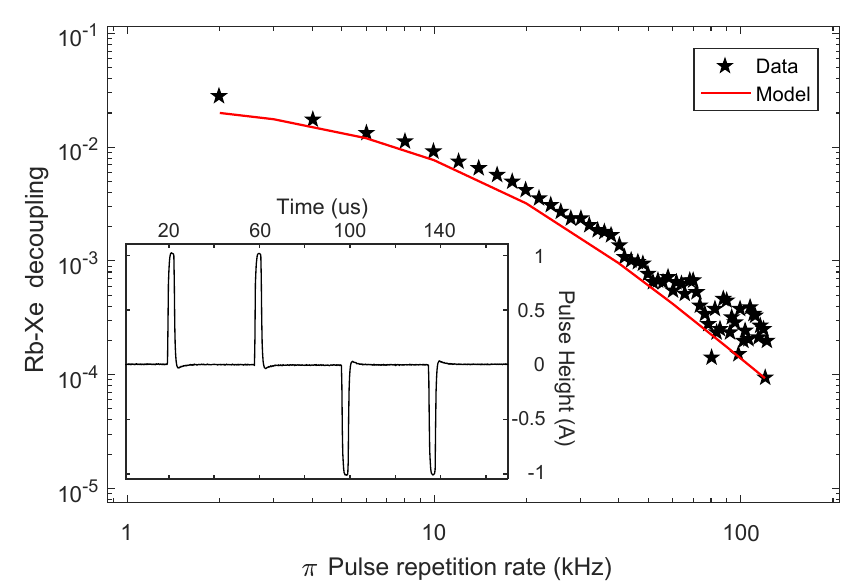}
\caption{The Rb back polarization from spin exchange with $^{129}$Xe  is reduced by a factor of $10^4$ by  increasing the pulse repetition rate, in agreement with predictions from a Rb density matrix model including Rb-Rb spin-exchange. The inset shows the current flowing through the coil to generate typical $\pi$ pulses.  }
\label{fig:2}
\end{figure}

In the presence of the noble gases, the pulse-train magnetometer provides additional advantages. It maintains zero average $^{87}$Rb polarization to minimize noble-gas frequency shifts. If the pump laser is turned off, it  actively depolarizes Rb atoms from back-polarization created by polarized $^{129}$Xe. Fig.~\ref{fig:2} shows the fraction of $^{87}$Rb back-polarization that remains in the presence of the pulses. This data was obtained by first polarizing $^{129}$Xe along the bias field, then shuttering the pump light and monitoring $^{87}$Rb back-polarization along the bias field using Faraday rotation, while periodically reversing $^{129}$Xe spins \cite{supp}. We realize a depolarization factor of 10$^{4}$,  about 2 orders of magnitude larger than was achieved using a continuous RF field resonant with $^{87}$Rb atoms  in \cite{Sheng_2014}.

{\em Noble-Gas Comagnetometer.--} The $^{3}$He-$^{129}$Xe comagnetometer uses a 0.5 cm$^3$ spherical cell made from GE180 glass, with 9.3 atm $^{3}$He, 2.9 torr $^{129}$Xe, 40 torr N$_2$, and a droplet of enriched $^{87}$Rb that plugs the cell stem. In this cell the magnetic linewidth is about 0.5 mG for optimized intensity of pump and probe lasers, due to additional broadening by Rb-Xe spin-destruction collisions. The magnetic field sensitivity of the pulse-train $^{87}$Rb magnetometer is 40 fT/$\sqrt{\text{Hz}}$ \cite{supp}. The nuclear-spin relaxation time due to wall collisions is about $10^3$ s for $^{129}$Xe and much longer for $^{3}$He. At the operating temperature of 110-120$^{\circ}$C the $^{129}$Xe spin relaxation time is about 200 s, limited by  Rb-Xe spin-exchange collisions. 

The comagnetometer is operated in a pump-probe cycle.  First $^{3}$He and $^{129}$Xe are polarized  along \(B_{0}\). We start the sequence for pulse-train $^{87}$Rb magnetometery and, using large field coils, apply a noble-gas\(\) tipping pulse  that places both $^{3}$He and $^{129}$Xe in the plane transverse to the bias field \cite{supp}. The tipping angle for $^{3}$He and $^{129}$Xe is crucial for accurate comagnetometry due to long-range dipolar fields generated by the noble gases in the imperfectly spherical cell \cite{supp}. However, dipolar fields do not cause a significant frequency shift if the nuclear-spin polarization is exactly orthogonal to $B_0$ and remains so throughout the measurement. It is important to tune the pump laser exactly to the D1 resonance, as the $^{87}$Rb pump light-shift can cause tipping of nuclear spins during the measurement. The precession signals from $^{3}$He and $^{129}$Xe are on the order of 0.3 mG; we limit the amplitudes to less than the magnetometer linewidth to avoid large non-linearities.  

\begin{figure} 
\center \includegraphics{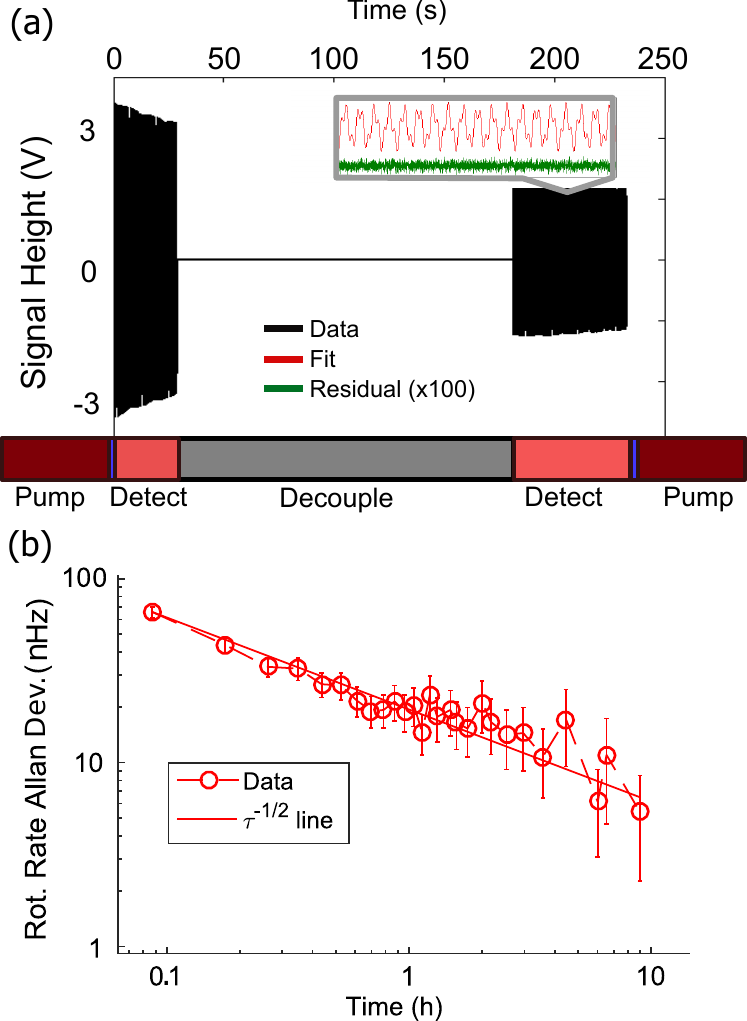}
\caption{(a) The Ramsey scheme with an inset showing a $^{3}$He-$^{129}$Xe lock-in signal pattern and the  fit residuals $\times100$. 
(b) Allan deviation of the rotation rate $\Omega_z$ with an 7 nHz ($10^{-2}$~deg/hour) upper limit of the rotation rate stability.}
\label{fig:3}
\end{figure}

The nuclear-spin precession measurements can be performed continuously with the pulse-train $^{87}$Rb magnetometer. However, we find the noble-gas long-term frequency stability is insufficient with this method because the $\hat{y}$ pulse train allows  back-polarization of $^{87}$Rb along the $\hat{y}$ axis \cite{supp}. Therefore, we apply two-axis depolarization using the pulse sequence
 $\pi_y\pi_x\pi_{y}\pi_{\text{-}x}\pi_{\text{-}y}\pi_x\pi_{\text{-}y}\pi_{\text{-}x}$. This prevents continuous use of the $^{87}$Rb magnetometer, so we use Ramsey-type `in-the-dark' measurements. After an initial precession period, the pump and probe lasers are turned off with AOMs and mechanically shuttered, and the  $\hat{x}$ $^{87}$Rb $\pi$ pulses are turned on, interspersing the $\hat{y}$ pulses.   
After waiting about $0.7~ T_{2}^\text{Xe}$, the pulse-train $^{87}$Rb magnetometer is used again to detect $^{3}$He-$^{129}$Xe precession, see Fig.~\ref{fig:3}a.
At the end of the precession measurement we  coherently put the remaining nuclear polarization along (or against) $B_0$ by sending the lock-in output to the $\hat{x}$ coil, which yields 
out-of-phase on-resonance magnetic fields for both  $^{3}$He and $^{129}$Xe \cite{Stockton_2004,Alem_2013}.  We then apply a magnetic field gradient to dephase any remaining transverse nuclear spin polarization.
The noble gases are then polarized along $B_0$ for about 20 s  using full Rb polarization, and the cycle repeats. 

We process the data by fitting each detection period to two decaying sine waves and extract the $^{3}$He and $^{129}$Xe zero-crossing times upon entering and exiting the dark period. We use the spin precession frequencies during the detection periods to find the integer number of precession periods between the zero-crossings in the dark. The $^{3}$He-$^{129}$Xe polarization signals are large enough that the $^{87}$Rb magnetometer response becomes slightly non-linear, causing $^{3}$He-$^{129}$Xe cross-modulation peaks. We correct for this effect by fitting with an expanded Lorentzian \cite{supp}.  From zero crossing times we  find the $^{3}$He and $^{129}$Xe `in-the-dark' precession frequencies $\omega_{\text{He}}$ and $\omega_{\text{Xe}}$, and the frequency ratio $f_r$. The rotation rate is given from Eq.~(\ref{eq:fr}),
\begin{equation}
 \Omega_z  = \omega_{\text{Xe}} \frac{  \gamma_r-f_r}{(\gamma_r -1)}, 
\end{equation}
where $\gamma_r = \gamma_{\text{He}}/\gamma_{\text{Xe}}$.
In Fig.~\ref{fig:3}b we show the Allan deviation of successive measurements of $\Omega_z$. The scatter for successive spin precession cycles is typically about 70 nHz and the fit indicates an angle-random-walk of 0.025~deg/$\sqrt{\text{hour}}$.

\begin{figure} 
\center
\includegraphics{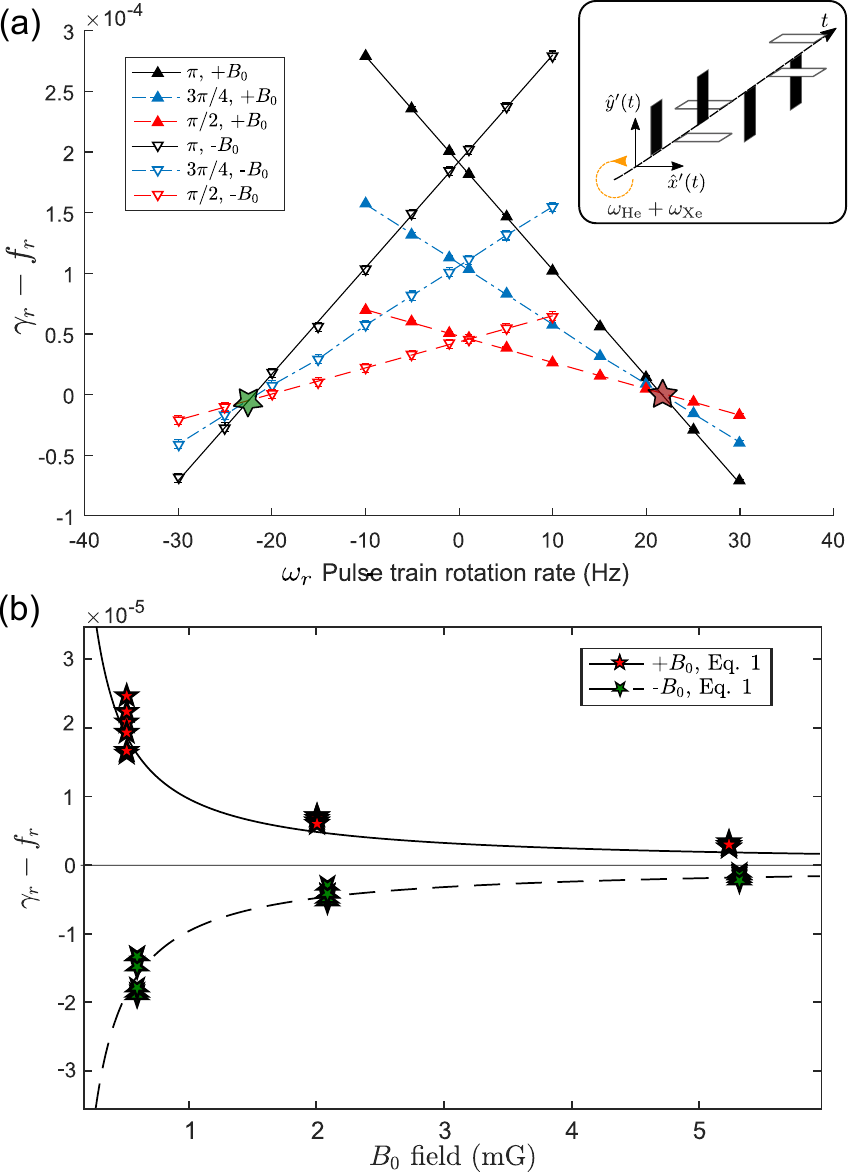}
\caption{(a) Frequency ratio $f_r\!=\!\omega_\text{He} /\omega_{\text{Xe}}$ measured with $^{87}$Rb depolarizing  $\hat{x}$-$\hat{y}$ pulse-train using  $\pi$, $3\pi/4$, and $\pi/2$ pulses and rotating at frequency $\omega_{r}$ for two directions of the bias magnetic field $B_0 = 5.3$ mG. The stars show the intersection points where the effect of the pulses is cancelled. 
Inset: rotating $\hat{x}$-$\hat{y}$ pulse train. 
(b) Frequency ratio $ \omega_\text{He} /\omega_{\text{Xe}}$ at the intersection points (note a factor of 10 expanded vertical scale) as a function of $B_0$. Solid and dashed lines show prediction from Eq.~\ref{eq:fr} due to projection of Earth's rotation on $B_0$.}
\label{fig:4}
\end{figure}

The last step in suppressing noble-gas frequency shifts is to eliminate the effect of $^{87}$Rb depolarizing pulses. To analyze the effects of the pulses on the average nuclear-spin precession frequency we use average Hamiltonian theory (AHT) \cite{supp}. The $\pi_y\pi_y\pi_{\text{-}y}\pi_{\text{-}y}$ $^{87}$Rb magnetometer sequence produces a nuclear-spin frequency shift $\omega_n= \gamma_n B_0 ( 1 - (\gamma_n B_1 t_p)^2/4)$, accurate to third order in $\gamma_n$ for sufficiently small pulse width $t_p$. The pulse field amplitude $B_1$ is given by $\gamma_\text{Rb} B_1 t_p=\pi$, so the frequency shift is a multiplicative correction  $(\gamma_n \pi/\gamma_\text{Rb})^2/4=5.3\times 10^{-5}$ for $^{3}$He and $7\times 10^{-6}$ for $^{129}$Xe frequencies. Finite pulse duration reduces the shifts by about 5\%.

To eliminate these effects we introduce a novel technique of slowly rotating the plane of the pulses. For example, consider rotating the pulse train at a rate $\omega_r$ that matches the $^{3}$He precession frequency---in the $^{3}$He rotating frame the $B_0$ field is eliminated and the $\pi$ pulses are the sole source of the magnetic field, so they produce no frequency shift if their time average is equal to zero. Experimentally this is achieved by applying current pulses to both $x$ and $y$ coils with amplitudes given by $\cos(\omega_r)$ and $\sin(\omega_r)$. This technique works for any shape of the current pulses and is insensitive to first-order inaccuracies in their relative amplitudes and phase. However, there remains a sensitivity to the \emph{planar} nature  of the pulses, for example, the presence of metal strips near the cell generates eddy currents that cause an additional apparent rotation of the pulse field. 

For the  $\pi_y\pi_x\pi_{y}\pi_{\text{-}x}\pi_{\text{-}y}\pi_x\pi_{\text{-}y}\pi_{\text{-}x}$ decoupling sequence applied during the in-the-dark period, AHT gives a frequency ratio of
\begin{equation}
f_r = \frac{\gamma_\text{He} B_0+ 3(\omega_r-\gamma_\text{He} B_0)(\gamma_{\text{He}} B_1 t_p)^2/8}{\gamma_{\text{Xe}} B_{0} + 3(\omega_r-\gamma_\text{Xe} B_0)(\gamma_{\text{Xe}} B_1 t_p)^2/8}.
\label{eq:AHT}
\end{equation}
We find that $f_r=\ \gamma_r$ for  $\omega_r = \omega_{\text{He}}+ \omega_{\text{Xe}}$ and any $B_1$. 

In Fig.~\ref{fig:4}a we show the measurements of the frequency ratio as a function of the rotation frequency $\omega_r$ and $B_1$ amplitude. We take data for both directions of $B_0$ and reverse the direction of pulse rotation. The rotation frequency can be set to one of  the intersection points, where $f_r$ is independent of the amplitude of $B_1$, eliminating the shift due to pulses. It corresponds to the sum $\omega_{\text{He}}+ \omega_{\text{Xe}}$, but can be slightly shifted if the pulses are not perfectly planar. 

To check the accuracy of the comagnetometer, we measure the difference $f_r-\gamma_r$ due to Earth's rotation. We use $\gamma_r=2.7540813(3) $ from \cite{Makulski_2015}.  The chemical shift of $^{129}$Xe frequency due to He is estimated to be on the order of 0.2 ppm in our cell \cite{Adrian}.  The data are shown in Fig.~\ref{fig:4}b as a function of $B_0$.  The projection of the  Earth's rotation axis $\mathbf{\Omega}_{\text{E}}$ along the bias field is $\Omega_z = \mathbf{\Omega}_{\text{E}}  \cdot \mathbf{B}_0/B_0= 6.5~\mu$Hz. Although $B_0$ is not parallel to the Earth's rotation axis, the effect of Berry's phase is negligible \cite{Suter_1987,Suter_1988,Appelt_1994,Appelt_1995}.
The predictions from Eq.~(\ref{eq:fr}) agree with our measurements without any frequency corrections, indicating absolute accuracy of frequency measurements at a level of about 1 ppm.
Allan deviation measurements in Fig.~\ref{fig:3} using the rotating decoupling field indicate that under stable experimental conditions the  frequency stability is better 7 nHz or about 1 ppb. This corresponds to an upper limit on angular bias drift of 0.009 deg/hour.

In conclusion, we have developed techniques for operating a dual noble-gas comagnetometer with isotopes that have very different values of $\kappa_0$ while using Rb vapor  in the cell for polarization and detection of nuclear spins.  We demonstrated a Rb magnetometery technique using  $\pi$ pulse trains that suppresses Rb-Rb spin-exchange relaxation in a finite magnetic field and maintains zero average Rb polarization. To further eliminate frequency shifts from polarized Rb we use a two-axis $\pi$-pulse train and make Ramsey-style measurements in the dark. By rotating these decoupling $\pi$-pulses we achieve sufficient absolute accuracy to  measure Earth's rotation without changing the comagnetometer orientation. We realize stability of 7 nHz after 8 hours, which is sufficient for detection of Planck-scale coupling of spins to Earth's gravity  \cite{Flambaum_2009}.  The comagnetometer can also be used as a gyroscope with higher sensitivity and stability than previous nuclear-spin gyroscopes that use a $^{129}$Xe-$^{131}$Xe combination with similar values of $\kappa_0$ \cite{Walker_2016,Korver_2015}. The short-term sensitivity of our apparatus using a 0.5 cm$^3$ cell is an order of magnitude better than for the $^3$He-$^{129}$Xe spin maser  \cite{Glenday_2008} and similar to the SQUID-detected  $^3$He-$^{129}$Xe comagnetometer that has a 300 times larger cell volume \cite{Tullney_2013}.  The best short-term sensitivity we realized experimentally is 0.01 deg/$\sqrt{\text{hour}}$,  while the Cram\'er-Rao frequency uncertainty lower bound based on the signal-to-noise ratio of the recorded $^3$He and $^{129}$Xe signals corresponds to an angle-random walk of 0.0002 deg/$\sqrt{\text{hour}}$. Better understanding of the Rb magnetometer cross-modulation peaks and short term frequency instabilities is necessary to realize the full sensitivity potential. 

This work was supported by DARPA and NSF. 

%

\end{document}


\begin{widetext}

\section{Supplemental material}
 
\subsection{Details of pulse-train $^{87}$Rb magnetometer}
In Fig.~\ref{fig:Transsig} we show the increase in the signal amplitude of the pulse-train magnetometer due to a \(B_{y}\) transverse magnetic field as the $\pi$ pulse rate is increased.
In the limit that the time between the pulses $\tau$ and the duration of the pulses $t_p$ satisfy the relationship  $t_p\ll \tau\ll 1/(\gamma_{\text{Rb}} B_0)$, the magnetometer signal is the same as for a SERF magnetometer operating at zero field. The magnetometer sensitivity and its dependence on the pump and probe beam intensities is also the same as for a SERF magnetometer. However, unlike a typical SERF magnetometer, the sensitive axis in this case is defined by the direction of the $\pi$ pulses along the $\hat{y}$ axis, not the laser directions, making it less sensitive to laser beam misalignment and presence of magnetic fields in the $\hat{x}$ and $\hat{z}$ directions. 

\begin{figure}[h!]
\center
\includegraphics[width = 3.25in]{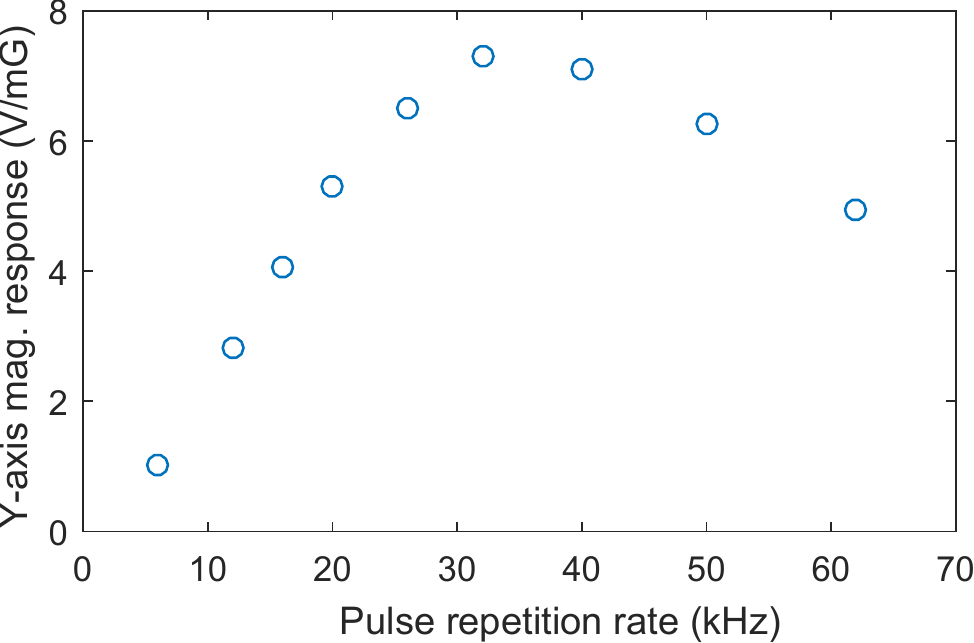}
\caption{The response of the $^{87}$Rb $\pi$ pulse magnetometer to a DC $B_{y}$ magnetic field as a function of the $\pi$ pulse repetition rate. The data correspond to the same conditions as Fig.~1 of the main paper, using a cell with N$_{2}$-only buffer gas and low laser intensity. }
\label{fig:Transsig}
\end{figure}

In Fig.~\ref{fig:Sens} we show the magnetic field signal and the sensitivity of the $^{87}$Rb magnetometer in the comagnetometer cell.  The presence of $^{129}$Xe broadens the Rb resonance linewidth, so the sensitivity is not as good as could be obtained in a typical magnetometer cell optimized to achieve the maximum $T_2$ for $^{87}$Rb.  We obtain a sensitivity  of 40~fT/Hz$^{1/2}$ in the comagnetometer cell. 

\begin{figure}[h!]
\center
\includegraphics[width = 7in]{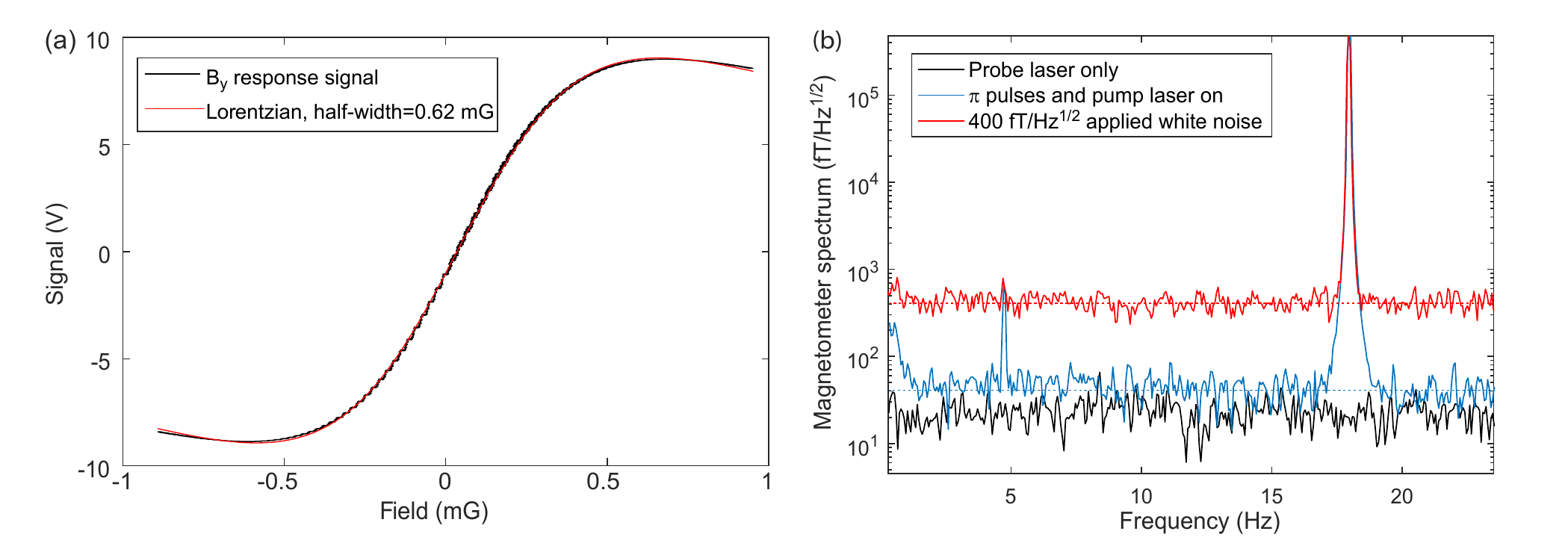}
\caption{(a) Magnetic field dispersion signal optimized for maximum magnetic field sensitivity in the comagnetometer cell. \\(b) The noise spectrum of the $^{87}$Rb magnetometer in the comagnetometer cell for 30 kHz $\pi$ pulse rate. }
\label{fig:Sens}
\end{figure}

In Fig.~\ref{fig:nonLin} we show the spectrum of the comagnetometer signal with large $^{3}$He and $^{129}$Xe spin precession signals at 17 Hz and 6 Hz respectively.  The signal-to-noise ratio is approximately $10^6$ and is limited by the dynamic range of the SR865 lock-in amplifier.  One can see a number of cross-modulation peaks due to non-linearity of the magnetometer for signal amplitudes comparable to the linewidth. We reduce the cross-modulation peaks by applying linearizing corrections proportional to the 3rd and 5th powers of the signal. We also apply a correction proportional to the derivative of the signal to account for the finite time response of the Rb magnetometer. However, we cannot completely account for all cross-modulation peaks. The signal amplitudes are optimized by changing the spin-exchange optical pumping time to achieve the smallest fitting uncertainties.

\begin{figure}[h!]
\center
\includegraphics[width = 3.25in]{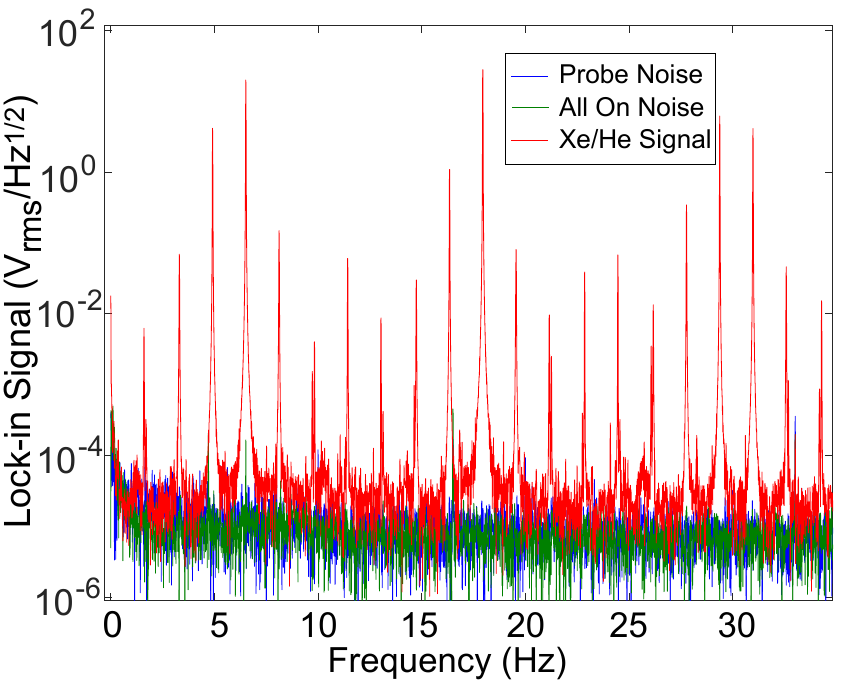}
\caption{Fourier spectrum of the $^{87}$Rb magnetometer signal with  $^{3}$He and $^{129}$Xe spin precession signals at 17 Hz and 6 Hz, respectively.  One can see a number of cross-modulation peaks at sums and differences of multiples of these frequencies because the signal amplitudes approach the linewidth of the Rb magnetometer.   }
\label{fig:nonLin}
\end{figure}

\subsection{Optimization of the comagnetometer operation}

To optimize the operation of the comagnetometer we measure the $^3$He to $^{129}$Xe frequency ratio $f_r$ as a function of various parameters.
In Fig. \ref{fig:PulseLengthSweep}, we show the ratio $f_r$ minus the nominal frequency ratio $\gamma_r$ as a function of the $^{87}$Rb $\pi$ pulse length.  
If the $^{87}$Rb $\pi$ pulses are imperfect during the detection periods and the $\sigma_+$-$\sigma_-$ pumping is not symmetric, then there exists a net $^{87}$Rb polarization along the $\hat{z}$ axis, which causes noble gas frequency ratio  shifts. While this problem is reduced  by using the Ramsey scheme with measurements in the dark, it still may cause $^{129}$Xe and $^3$He polarization to build up along $\hat{z}$, which causes additional frequency shifts as described below.  The $^{87}$Rb magnetometer signal is maximized when  $\sigma_+$-$\sigma_-$ pumping is symmetric and $\pi$ pulses are accurate, which allows for rough tuning of the system. Additionally, for accurate $^{87}$Rb $\pi$ pulses, the frequency ratio during the detection period becomes independent of the pump intensity and direction of the bias magnetic field. Flipping the magnetic field is equivalent to flipping the phase of the EOM modulation that generates the $\sigma_+$-$\sigma_-$ pumping light. This allows one to also fine-tune the EOM voltage bias to obtain $\sigma_+$-$\sigma_-$ symmetry.

\begin{figure}[h!]
\center
\includegraphics[width = 3.25in]{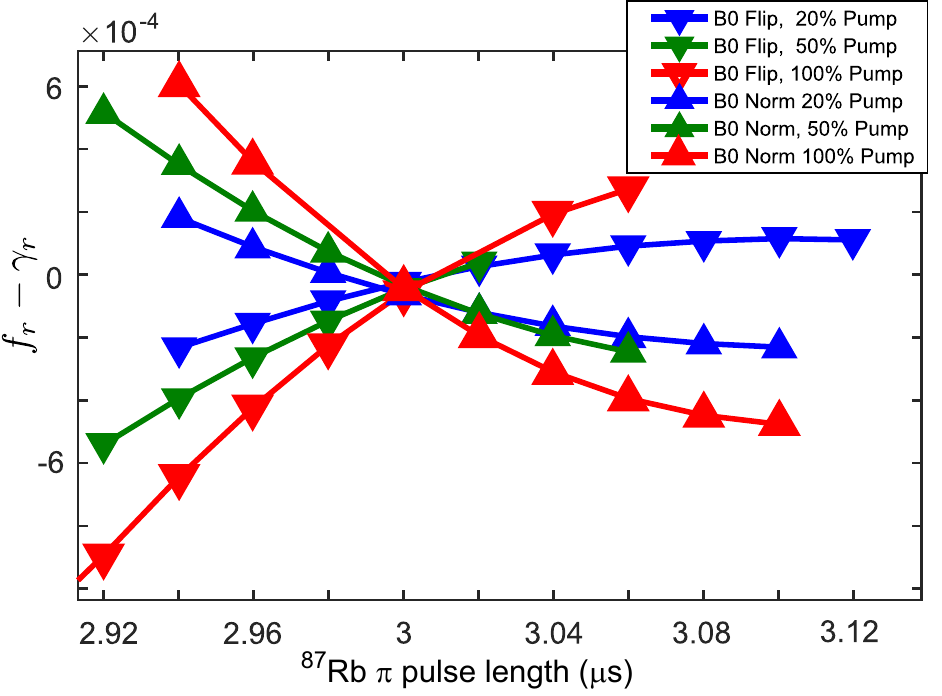}
\caption{The effect of inaccurate $^{87}$Rb $\pi$ pulses on the frequency ratio $f_r$ at standard pumping conditions is shown. We sweep through  $\pi$ pulse lengths for different optical pumping powers, then flip the field and perform the same sweep.  }
\label{fig:PulseLengthSweep}
\end{figure}

The frequency ratio measured in-the-dark is sensitive to small $^{129}$Xe and $^3$He polarizations along the bias field. This effect for $^{129}$Xe is seen in Fig.~\ref{fig:XYvY}, where we purposefully vary the $\pi/2$ $^{129}$Xe tipping pulse. The mechanism of this for $\hat{y}$-only $\pi$ pulses is as follows. When the $^{129}$Xe polarization points in the $\hat{y}$ direction, a small Rb polarization develops due to spin exchange with  $^{129}$Xe. The Rb polarization along $\hat{y}$ is not averaged by the $\pi$ pulses along  $\hat{y}$.  This Rb polarization causes an effective field, felt much more strongly by  $^{129}$Xe than by  $^{3}$He, because of the larger value of $\kappa_0$. The effective field oscillates along the $\hat{y}$ direction at the $^{129}$Xe precession frequency. The oscillating field can be decomposed into two rotating components, one of which co-rotates with $^{129}$Xe transverse polarization. If there is any $^{129}$Xe polarization along $B_0$, the effective field it will cause it to precess into the transverse plane, but out of phase with respect to main $^{129}$Xe signal. This causes a $^{129}$Xe frequency shift which can be estimated as follows:
\begin{equation}
\delta \omega=\frac{P^{\text{Xe}}_y [\text{Xe}] \left\langle \sigma_{\text{SE}}v\right\rangle_{\text{RbXe}}}{\Gamma_{\text{Rb}}} \frac{8 \pi\kappa_0 \mu_B [\text{Rb}]}{3} \frac{\gamma_{\text{Xe}}P^{\text{Xe}}_z}{2P^{\text{Xe}}_y} \label{Xefsh}
\end{equation} 
where the first fraction gives the Rb polarization along $\hat{y}$ due to spin-exchange with $^{129}$Xe, the second fraction gives the Rb magnetic field felt by Xe along $\hat{y}$; and the last fraction gives the frequency shift due to Xe polarization along $\hat{z}$ rotating into the transverse plane. 
Here $\kappa_0$ is the Fermi-contact enhancement factor, $\sigma_{\text{SE}}$ is the spin-exchange cross section, $v$ is the average relative velocity for Rb-Xe, $\Gamma_{\text{Rb}}$ is the relaxation rate of Rb, $P^{\text{Xe}}$ is the polarization of Xe along a given axis, $[\text{Xe}]$ and $[\text{Rb}]$ are the Xe and Rb number densities, $\gamma_{\text{Xe}}$ is $^{129}$Xe gyromagnetic ratio, and $\mu_B$ is the Bohr magneton. 
If the depolarizing pulses are applied along both  $\hat{x}$ and $\hat{y}$ directions, then Rb polarization does not build up along any axis and the frequency ratio becomes much less sensitive to the absolute value of the $^{129}$Xe polarization and to the accuracy of the tipping pulses. The remaining shift in the frequency ratio seen in Fig.~\ref{fig:XYvY} is much less than $10^{-6}$ under our typical operating conditions when the Xe amplitude is about 2 V. 
\begin{figure}[h!]
\center
\includegraphics[width = 3.25in]{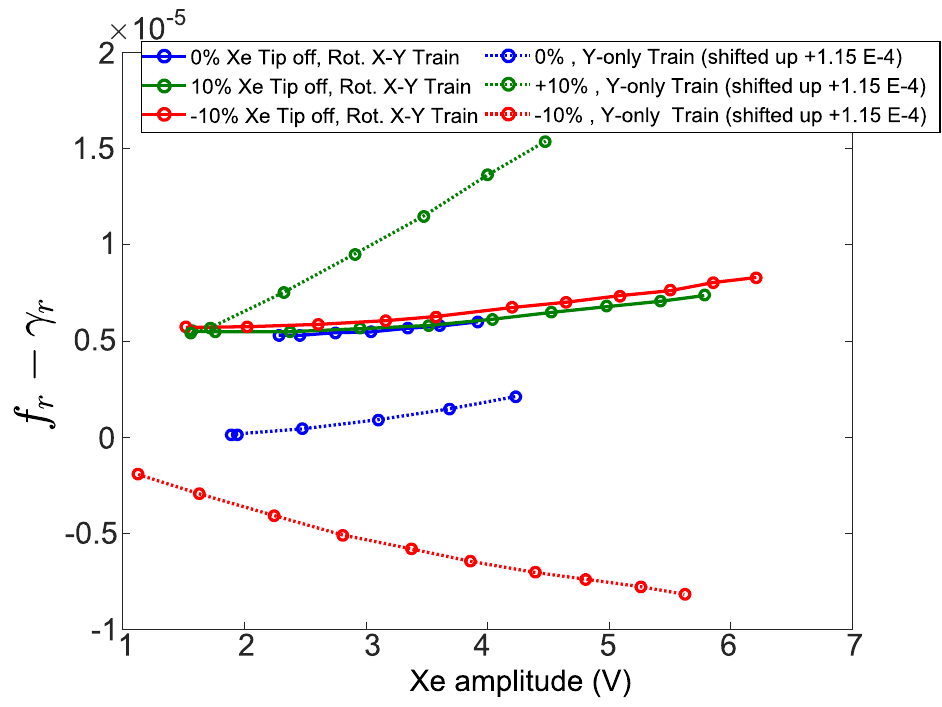}
\caption{ Frequency ratio $f_r$ measured in the dark as a function of the $^{129}$Xe polarization for $^{129}$Xe $\pi/2$ tipping pulses that are -10\%, 0\% , and 10\% off from ideal.  With  $\hat{y}$ pulses in the dark, there is a significant change in the frequency ratio $f_r$ that is dependent on $^{129}$Xe polarization due to back-polarization of Rb.  For $\hat{x}$-$\hat{y}$ depolarizing pulses, the frequency ratio shift is much smaller and is insensitive to the accuracy of the $\pi/2$ pulse.}
\label{fig:XYvY}
\end{figure}	

During the detection intervals we apply only $\hat{y}$ $\pi$ pulses since it is impossible to operate a magnetometer while simultaneously depolarizing Rb along all 3 axes. Additional complications can arise if the pump laser is not exactly tuned to the $^{87}$Rb optical resonance since it causes a light shift, which reverses sign with $\sigma_+$-$\sigma_-$ modulation. Such a modulated field along $\hat{z}$ direction is not decoupled by the $\pi$ pulses. For example, consider pulsed magnetometer operation with a $\hat{y}$  $^{129}$Xe magnetic field signal, the Rb polarization has a component that alternates between $+\hat{x}$ and $-\hat{x}$ directions with every $\pi$ pulse. The Rb spin precession around a constant  $\hat{z}$ bias field is canceled by the  $\pi$ pulses. However, if the  $\hat{z}$ field has a component that is also modulated with the $\pi$ pulses, it will cause a net precession of the Rb $\hat{x}$ polarization into $\hat{y}$ direction. This effect can be much larger than Rb back-polarization by spin-exchange with $^{129}$Xe, since the rate of spin precession due to the lightshift can approach the optical pumping rate and the linewidth of the Rb magnetometer.  Experimentally, we have exaggerated this effect by tuning the Rb pump laser 60 GHz away from resonance (comparable to the optical resonance linewidth) and detuning the $^{129}$Xe pulses by $15\%$ from $\pi/2$, then the frequency ratio $f_r$ drift during a single shot (200 s) is on the order of $3\times10^{-4}$, about an order of magnitude larger than shown in Fig. \ref{fig:XYvY} for the Rb spin-exchange back-polarization. Even if the  $^{129}$Xe $\pi/2$ pulses are perfectly accurate, a second-order effect due to the light shift can cause $^{129}$Xe to develop a $\hat{z}$ polarization over time.  Any Rb polarization in the $\hat{y}$ direction will precess around  $\hat{z}$ field. If the $\hat{z}$ field is itself modulated by the light shift, there will be a net time-averaged Rb polarization in the $\hat{x}$ direction, as it will not be perfectly averaged between  $\hat{x}$ and $-\hat{x}$. Then the net Rb polarization in the $\hat{x}$  direction will cause a rotation of the  $^{129}$Xe polarization from $\hat{y}$ to $\hat{z}$ axis. For these reasons we find that the comagnetometer performs about an order of magnitude better when using frequency measurements in the dark and employing  $\hat{x}$ and $\hat{y}$ depolarizing pulses.

We observe a small rise in signal during the second detection period of the Ramsey sequence, which we have correlated with cooling of the cell from laser being blocked during the dark period. During the second detection period, the cell is laser heated again and the $^{87}$Rb density slowly (1-10 s) increases to steady state. We are able to include this into our fitting routine by adjusting the fitted signal amplitude. We do not find any evidence of a frequency ratio shift associated with this temperature change.

Another source of frequency instability is the classical dipolar fields produced by noble gas magnetization. This effect exists in all noble gas comagnetometers to some extent. The two noble gases generate comparable signals in our $^{87}$Rb magnetometer, but since $\kappa_0$ is much smaller for $^3$He, it corresponds to a larger $^3$He magnetization, typically on the order of 5 $\mu$G. The dipolar fields from $^3$He are further suppressed because we use a spherical cell and plug its stem with a drop of $^{87}$Rb atoms. However, small imperfections in the cell still generate a residual dipolar field. When  $^3$He is precessing around $B_0$ there will be a transverse dipolar field that rotates or oscillates at the  $^3$He frequency, depending on the exact shape of the cell. The primary effect of this field is to tip any  $^3$He polarization initially along $B_0$ into the transverse plane. The  $^3$He frequency shift can be calculated similar to Eq.~\ref{Xefsh},
\begin{equation}
\delta \omega=\frac{d \mu_{\text{He}} [\text{He}] P^{\text{He}}_{y}  \gamma_{\text{He}} P^{\text{He}}_{z}}{P^{\text{He}}_y}
\end{equation}
where $d$ is a dimensionless coefficient describing the strength of the rotating dipolar field relative to the $^3$He magnetization  $M_{\text{He}}=\mu_{\text{He}} [\text{He}] P^{\text{He}}_{y}$. The dipolar field slowly rotates the $P^{\text{He}}_{z}$ into the transverse plane out of phase relative to $P^{\text{He}}_{y}$, causing a frequency shift. A static dipolar field along $B_0$ will also shift $^3$He and $^{129}$Xe frequencies, but they will shifted by the same amount, so the effect cancels in the frequency ratio. 
\begin{figure}[h!]
\center
\includegraphics[width = 3.25in]{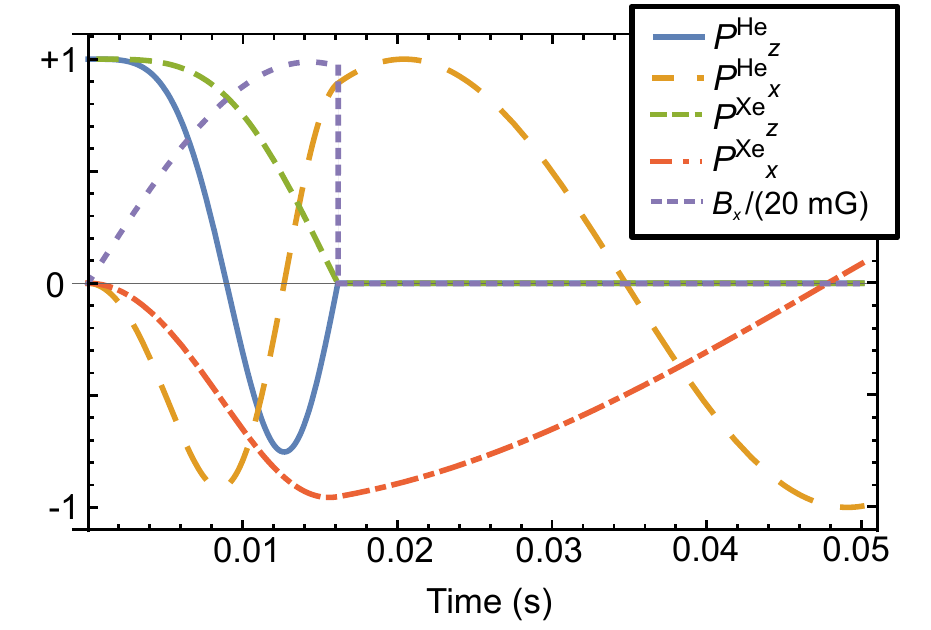}
\caption{An applied $B_x$ field with a 19.7 mG amplitude is shown, along with the responses of the projections of the $^{3}$He and $^{129}$Xe polarization on the $\hat{z}$ and $\hat{x}$ axes. Both $\hat{z}$ projections of the noble gases are zeroed at the end of the $B_x$ tipping pulse. }
\label{fig:CalcPlot}
\end{figure}

To place the $^{3}$He and $^{129}$Xe onto the plane transverse to the 5 mG bias field along $\hat{z}$, we first numerically solve for a $B_x(t)$ that, because $\gamma_{\text{He}}/\gamma_{\text{Xe}} \approx 2.75$, roughly creates a $\pi/2$ pulse for $^{129}$Xe and $3\pi/2$ pulse for $^{3}$He. 
We choose to keep the frequency of the pulse fixed at the $^{3}$He Larmor frequency, and numerically solve a system of ODEs for the duration $t_f$ and height $B_x$ that causes $P_z^{\text{He}}(t_f) = P_x^{\text{Xe}}(t_f) = 0$,  
Note we also choose to use a tipping field large enough to be well outside of the RWA approximation, so that numerical solutions to the non-linear problem are  necessary; we consider a pulse within the RWA to be too long in duration.  
In Fig.~\ref{fig:CalcPlot}, we show the results of such a calculation, demonstrating the effect of a $B_x$ tipping pulse on the projections of the $^{3}$He and $^{129}$Xe polarizations along the $\hat{z}$ and $\hat{x}$ axes. 

We show the effect of $^{3}$He $\pi/2$ pulses on our `in-the-dark' the comagnetometer measurement of $f_r$, in Fig.~\ref{fig:heTipper}.
This data is taken by initially placing the $^{3}$He polarization along or against the bias field $B_0$ (shown with + or - He amplitude), then varying the $\pi/2$ tipping pulse height. The crossing point of these measurements indicate the precise value of the $\pi/2$ tipping pulse height. The stability of $f_r$ at this point can be further verified by  changing the magnitude of the $^3$He polarization, although this a slow process because of slow Rb-$^3$He spin-exchange. 
\begin{figure}[h!]
\center
\includegraphics[width = 3.25in]{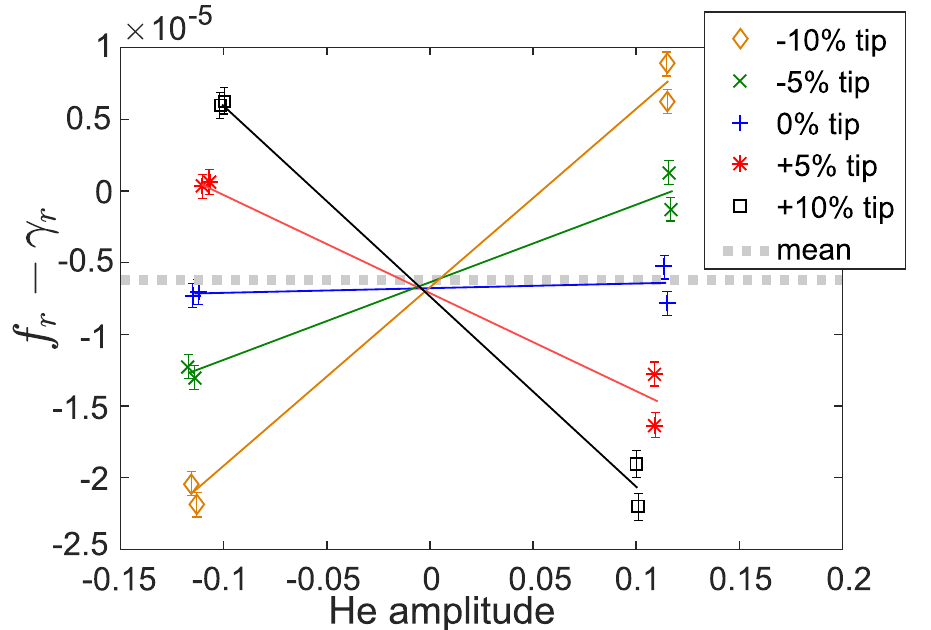}
\caption{Frequency ratio for  $^{3}$He tipping pulses that are -10\%, -5\%, 0\% , 5\%, and 10\% off from ideal and for both directions of initial $^{3}$He polarization. The frequency shift is caused by dipolar magnetic fields in an imperfectly spherical cell. }
\label{fig:heTipper}
\end{figure}

From the sensitivity of the frequency ratio to the $^{3}$He tipping pulse amplitude, we estimate the size of the residual dipolar field created by $^3$He due to imperfections of the cell to be on the order of 0.5 $\mu$G. This is roughly 1\% of the classical dipolar field created by uniform magnetization \cite{Romalis_2014}. We can also estimate the second-order Bloch-Siegert shift created by the counter-rotating component of the dipolar field to be on the order of $(0.5 \mu$G$/5$mG$)^2=10^{-8}$ fractional shift for $^3$He. The frequency shift for $^{129}$Xe due to rotating or oscillating off-resonance $^3$He field is on the same order of magnitude. 

\subsection{Technique for obtaining Rb-Xe decoupling data}
\begin{figure}[h!]
\center
\includegraphics[width = 3.25in]{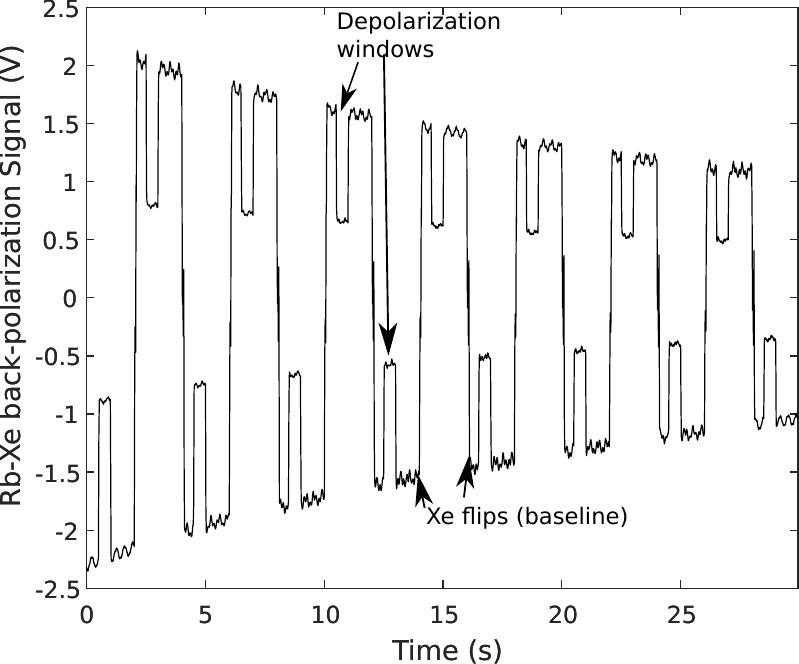}
\caption{Measurement of Rb back-polarization by $^{129}$Xe. The measurements are done in the absence of pump light by periodically reversing  $^{129}$Xe polarization and turning on and off the depolarizing pulses at a given rate. (Note the ripples in the signal are due to remnant $^{3}$He.)  }
\label{fig:supp}
\end{figure}
We have shown that back polarization of Rb by  Rb-Xe spin-exchange can be decoupled by a factor of $10^4$  (comparable to the hyperfine decoupling of K-Xe shown in Ref.~\cite{Sheng_2014}).
This data was taken by first hyperpolarizing $^{129}$Xe along a bias field.  We then turn off the pumping light, and use a Faraday modulator (14 kHz, fed into lock-in) to watch the paramagnetic Faraday rotation of the $^{87}$Rb due to the $^{129}$Xe backpolarizing the  $^{87}$Rb along the bias field. We then apply our decoupling pulses at a specific rate in our depolarization windows. To prevent any inaccuracies due to background optical rotation or pumping of  Rb  atoms by residual circular polarization of the probe laser, we quickly flip the $^{129}$Xe back and forth using $\pi$ pulses. 

There does not appear to be a fundamental limit on the degree of decoupling using this technique; in principle the pulse repetition rate can be increased and pulse width decreased for the desired decoupling.  To model the decoupling we include Rb-Rb spin-exchange, which modifies the slope of the decoupling curve around 20 kHz.

\subsection{Pulse Design and Average Hamiltonian Theory}
At high magnetic fields, decoupling techniques can be analyzed using Average Hamiltonian Theory (AHT), where the main focus is usually decoupling with respect to the bias field axis.
The appeal of AHT is that all higher-order correction terms are themselves Hermitian. 
Haeberlen and Waugh \cite{Haeberlen_1968} developed the technique to treat schemes in solid NMR---this treatment is essentially the use of the Magnus expansion, which is the continuous form of the Baker-Campbell-Hausdorff (BCH) series. 
Pines and coworkers \cite{Lee_1987} considered zero- and low-field regimes of Average Hamiltonian Theory and showed the importance of averaging along all three spatial axes. 
In particular, for effective decoupling between two coupled spins, there are three linear terms and six bilinear terms to average. 

If one spin species is affected by the pulse sequence much more strongly than others, the requirement at low field must be to average the three linear terms of its magnetization sufficiently fast. 
Using the notation of Ref.~\cite{Lee_1987}, we note there cannot be a perfect inversion of a vector due to a rotation, $XYZ \not\rightarrow \bar{X}\bar{Y}\bar{Z}$. 
Thus, we must rely on the rapid averaging of the vector, cycling between states such as $XYZ, X\bar{Y}\bar{Z}, \bar{X}\bar{Y}Z, \bar{X}Y\bar{Z}$, for example, using a sequence $\pi_x, \pi_y, \pi_x,\pi_y$, with a symmetric interval between each pulse. 
The application of these types of pulses with the purpose of decoupling $^{87}$Rb from noble gas spins requires some additional constraints:
(1) The path of $^{87}$Rb during the pulse must be considered, and should be averaged to minimize the Fermi-contact shift, e.g., a $\pi_y, \pi_{-y}$ sequence should not be used during detection, because it will introduce a large time-averaged Rb magnetization along the $x$-axis. 
(2) A helicity-free, symmetric pulse that minimally affects the noble gas precession should be used to mitigate large shifts due to the pulses themselves. 
Thus, $\pi_x, \pi_y, \pi_x,\pi_y$ should not be used, but rather, we use $\pi_y, \pi_x, \pi_y, \pi_{-x}, \pi_{-y}, \pi_x, \pi_{-y}, \pi_{-x}$. 
Note that the latter pulse gives a counter-clockwise, then clockwise sense when looking down the $z$-axis. 
Symmetry also implies perfect averaging of the $^{87}$Rb polarization vector if and only if there is no phase accumulation by the $^{87}$Rb. 
This means that perfect averaging along all 3 axes is not possible if we want to use a Rb magnetometer where the signal grows on a time scale much longer than the pulse repetition rate. 
 
If we assume that our pulses are sufficiently fast and effectively decouple alkali-noble gas  spin-exchange effects (see Fig.~2 in main text), we can rely on the symmetry of our pulses to analyze how much the pulses themselves affect the precession frequency of the noble gases within a piece-wise treatment of AHT. 
When using a piece-wise treatment, the problem boils down to using the Baker-Campbell-Hausdorff (BCH) relation to approximate the solution to 
\begin{equation}
e^{Y} = \prod_{j = i}^{1} e^{X_j}.       
\end{equation}
This can be solved by repeatedly applying BCH and collecting terms of a certain order, but we have found it much quicker computationally to solve for the order of terms in form of a summation. 
We first consider a bias field, four single-axis pulses  ($\pi_y, \pi_y, \pi_{-y}, \pi_{-y}$), then show the effect of rotating these pulses. 
With $\gamma$ as the nuclear gyromagnetic ratio, $t_c$ as the cycle time, $\tau$ is the pulse repetition time, $t_p$ as the pulse time, $B_0$ as the static bias field, and $B_1$ as the pulse height, we check up to the fourth order AHT corrections (only up to third order are presented here)
\begin{equation}
\begin{split}
\bar{H}_0 t_c &=\gamma B_0 I_z t_c\\
\bar{H}_1 t_c &= -i \gamma^2 B_0 B_1 I_x t_p t_c\\
\bar{H}_2 t_c &= \frac{1}{3} \gamma^3 B_0 B_1 t_p (B_1 I_z t_p (t_p - 9 \tau) + 6 B_0 I_y (t_p - \tau) \tau)\\
\bar{H}_3 t_c &=  \frac{1}{3} i\gamma^4 B_0 B_1 I_x t_p (B_0^2 \tau (2 t_p^2 - 3 t_p \tau - 3 \tau^2)-B_1^2 t_p^2 (t_p - 5 \tau)).
\end{split}
\end{equation}
Note that in the limit of $t_p/t_c \rightarrow 0$, the full Hamiltionian is given by $\bar{H_0}$.

Collecting the terms that dominate the $\gamma^3$ order in the precession frequency requires we collect $\bar{H}_i$ up to second order, 
\begin{equation}
\bar{H}_{0\text{-}2} = I_z (\gamma B_0- \gamma^3 \frac{3}{4} B_0 B_1^2 t_p^2(1-\frac{t_p}{9\tau}) )   + I_x \gamma^2 B_0 B_1  t_p, 
\end{equation}
which, for sufficiently small $t_p$ causes a precession frequency
\begin{equation}
\omega_{\text{}} \approx \gamma B_0 \left( 1 - (\gamma B_1 t_p/2)^2\right) .
\label{eq:4pRotation}
\end{equation}
Since we require that a $^{87}$Rb $\pi$-pulse is made, we can substitute $t_p = \pi/(\gamma_{Rb} \sqrt{B_1^2 + B_0^2})\approx \pi/\gamma_{Rb} B_1$. 

Now, we consider the effect of rotating these pulses with a frequency $\omega_{r}$. 
This is greatly simplified by considering a rotating frame that is moving at the frequency of the pulse train rotation, where the problem becomes identical to what yielded us the static solution given in Eq.~\ref{eq:4pRotation}, with the substitution $\gamma B_0 \rightarrow \gamma B_ 0 - \omega_{r}$. 
This is because the rotating pulses ($\pi_{y'}, \pi_{y'}, \pi_{-y'}, \pi_{-y'}$) are static in this rotating frame, leaving the AHT/BCH problem identical to the one solved above, with the effective bias field transformed similar to a conventional rotating-frame description. 
Writing the noble gas frequency back in the laboratory frame, we have 

\begin{equation}
\omega= \omega _{r} + (\gamma B_0 - \omega_{r})  \left( 1 - (\gamma B_1 t_p/2)^2\right)
\end{equation}

Similarly, collecting dominating terms for our in-the-dark decoupling sequence $\pi_y, \pi_x, \pi_y, \pi_{-x}, \pi_{-y}, \pi_x, \pi_{-y}, \pi_{-x}$ results in the first order term, $\bar{H}_1 t_c = -i \gamma^2 B_0 B_1 t_p t_c (2I_x - I_y)/2$, and relevant part of the second order term, $\bar{H}_2 = -\gamma^3 B_0 B_1^2 t_p^2  I_z (1 - t_p /12\tau)$.
We find
\begin{equation}
\omega= \omega _{r} + (\gamma B_0 - \omega_{r}) \left( 1 - \frac{3}{8}(\gamma B_1 t_p)^2\right).
\label{eq:AHT}
\end{equation}

For rotating pulses at frequency $\omega_r$ {\em and rotating apparatus} at frequency $\Omega$, the effect on a single species becomes
\begin{equation}
\omega = \gamma B_0 +\Omega- (\gamma B_0 - \Omega - \omega _{r})\left(\frac{3 (\gamma B_1 t_p)^2}{8}\right),
\end{equation}
so that the effect of the pulses are technically only zeroed when $\omega_{r}= \gamma B_0 - \Omega$. 
However, since in our experiments $\Omega \ll \gamma_{\text{Xe}}B_0$ and the pulses are rotated at $\omega = \omega_{\text{He}}+ \omega_{\text{Xe}}$, Eq.~\ref{eq:AHT} is sufficiently accurate.

\end{widetext}
%